\documentclass[aps,pra,twocolumn,superscriptaddress,amsmath,amssymb,letterpaper]{revtex4}

\ifx\pdfoutput\@undefined\usepackage[usenames,dvips]{color}
\else\usepackage[usenames,dvipsnames]{color}
\usepackage{graphicx}
\usepackage{bm}
\usepackage{amssymb}
\usepackage{hyperref}
\usepackage{color}

\usepackage{amsmath}

\newcommand{\be}{\begin{equation}}
\newcommand{\ee}{\end{equation}}

\newcommand{\bs}{\boldsymbol}

\begin{document}

\title{Acoustic vortex beams in synthetic magnetic fields}

\date{\today}

\author{Irving Rond\'on}
\affiliation{Center for Theoretical Physics of Complex Systems, Institute for Basic Science (IBS), Daejeon 34126, Republic of Korea
}

\author{Daniel Leykam}
\affiliation{Center for Theoretical Physics of Complex Systems, Institute for Basic Science (IBS), Daejeon 34126, Republic of Korea
}

\begin{abstract}
We analyze propagation of acoustic vortex beams in longitudinal synthetic magnetic fields. We show how to generate two field configurations using a fluid contained in circulating cylinders: a uniform synthetic magnetic field hosting Laguerre-Gauss modes, and an Aharonov-Bohm flux tube hosting Bessel beams. For non-paraxial beams we find qualitative differences from the well-studied case of electron vortex beams in magnetic fields, arising due to the vectorial nature of the acoustic wave's velocity field. In particular, the pressure and velocity components of the acoustic wave can be individually sensitive to the relative sign of the beam orbital angular momentum and the magnetic field. Our findings illustrate how analogies between optical, electron, and acoustic vortex beams can break down in the presence of external vector potentials.
\end{abstract}

\maketitle

\section{Introduction}

Analogies between wave phenomena in different physical systems are a powerful tool to understand and control wave propagation. For example, the idea of characterising electronic Bloch waves using topological invariants, responsible for the discovery of a wealth of new topological insulator materials~\cite{topological_insulator_review}, has been fruitfully translated to photonics~\cite{topological_photonics_review} and acoustics~\cite{topo_acoustics,ge2018,berry_acoustic,topo_acoustic_review}. Similarly, wave orbital angular momentum, long-studied in the context of optical vortex beams~\cite{OAM_book}, can also be created for free electron beams in transmission electron microscopes~\cite{electron_review}. Going beyond the simplest scalar waves, spin and spin-orbit-interactions provide extra degrees of freedom to design and control structured vectorial wave fields~\cite{spin_review}.

Acoustic waves are longitudinal and conventionally viewed as spinless, purely scalar waves. Recently, however, several studies have found that acoustic waves can exhibit nontrivial vectorial properties in the form of spin textures and spin-momentum locking~\cite{Ylong,Shi_arxiv,Bliokh_arxiv,bliokh2019a,bliokh2019b}. These features are hidden in the commonly-used second order Schr\"odinger-like wave equation governing the acoustic velocity potential $\phi$ and require analysis of the more fundamental first order equations describing the coupled acoustic pressure $P$ and velocity ${\bf v}$ fields. While the acoustic spin necessarily vanishes for the simplest plane waves, nontrivial spin-related phenomena can emerge for localized fields such as surface waves and non-paraxial beams~\cite{Shi_arxiv,Bliokh_arxiv,bliokh2019a}. Spin therefore provides an additional degree of freedom to control acoustic waves, with potential applications including particle manipulation~\cite{acoustic_vortex}, selective excitation of surface waves~\cite{Shi_arxiv}, and design of novel topological materials. For example, Ref.~\cite{Shi_arxiv} recently observed the spin-controlled directional excitation of acoustic waves at the surface of a metamaterial waveguide.

Since spin is a form of angular momentum, it is natural to ask how acoustic spin interacts with the more widely-studied orbital angular momentum degree of freedom, exemplified by vortex beams~\cite{Lekner,bliokh2010,Zhang,acoustic_splitter}. Nontrivial spin in acoustic vortex beams was already hinted at by Zhang and Marston in 2011~\cite{Zhang}, who showed that the full acoustic angular momentum flux tensor contains nonparaxial corrections. Bliokh and Nori recently employed a quantum-like formalism analogous to that used for electromagnetic waves to show that the angular momentum of acoustic beams naturally divides into spin and orbital parts, obtaining nonzero spin densities of acoustic Bessel beams in free space~\cite{bliokh2019a}. Nevertheless, the total spin always vanishes, as required for longitudinal waves. Due to the time-reversal symmetry of the underlying acoustic wave equations, reversing the orbital angular momentum of an acoustic vortex beam must flip its spin density. 

\begin{figure}

\includegraphics[width=\columnwidth]{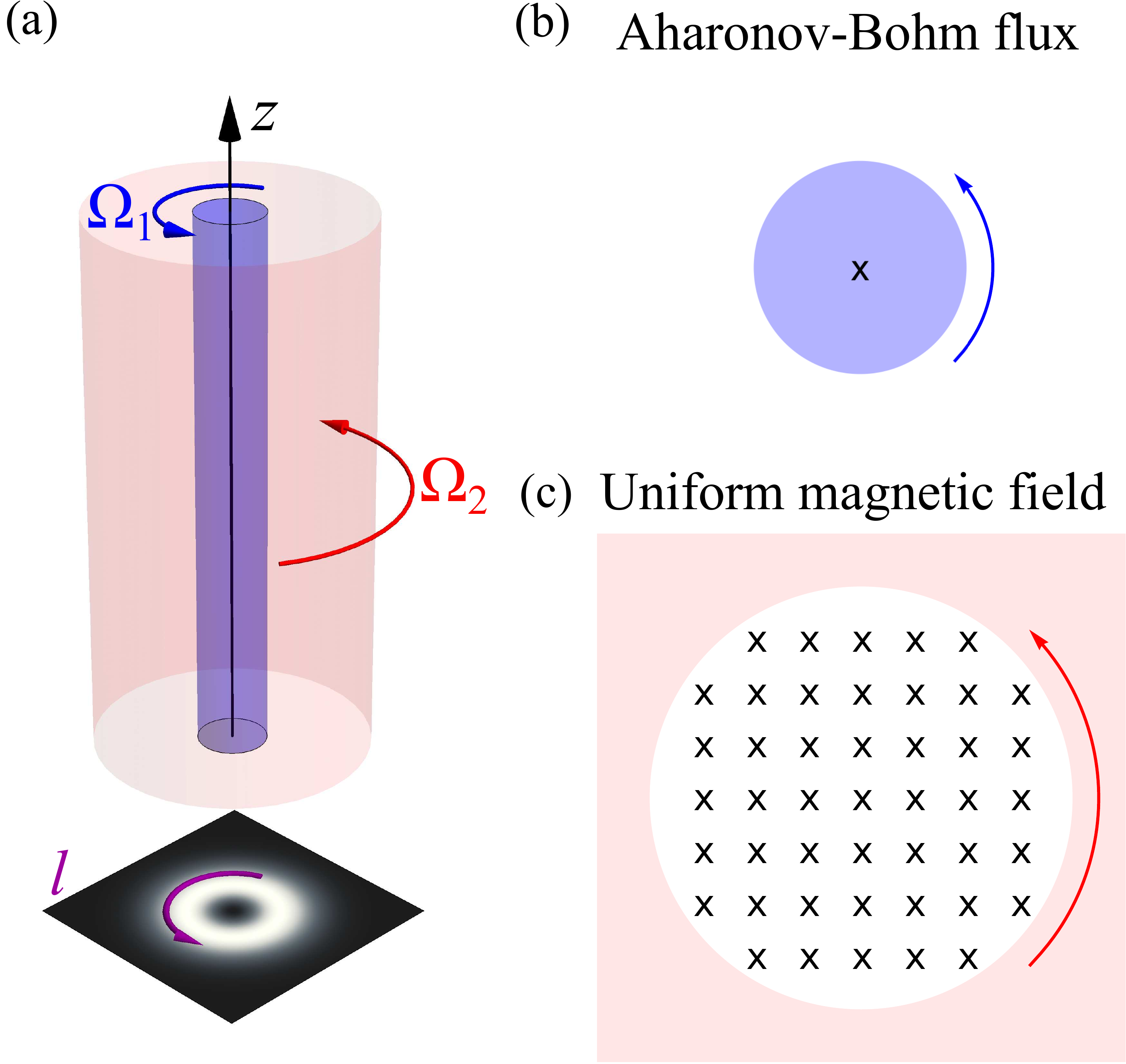}

\caption{Schematic of the studied system. (a) Propagation of a charge $l$ acoustic vortex beam with frequency $\omega$ through a waveguide formed by concentric cylinders with radii $R_{1,2}$ rotating with angular frequencies $\Omega_{1,2}$. (b) Inner rotating cylinder creates an Aharonov-Bohm tube of flux $\alpha = \omega \Omega_1 R_1^2 / c^2$, where $c$ is the speed of sound. (c) The outer (hollow) cylinder creates a uniform synthetic magnetic field of strength $\Omega_2$.}
\label{fig:system}

\end{figure}

In this manuscript we aim to complement these recent studies by analyzing the effect of broken time-reversal symmetry on the spin, momentum, and energy densities of acoustic beams. To break time-reversal symmetry we consider acoustic waves propagating through a slowly circulating fluid, which generates synthetic vector potentials and magnetic fields for sound. Previous studies focused on the quasi-2D limit of acoustic waves propagating in the plane of circulation, studying scattering off a magnetic flux tube~\cite{berry}, acoustic isolation and nonreciprocal transmission~\cite{fleury2014}, and topological phononic crystals~\cite{topo_acoustics}. In these examples, the synthetic magnetic field is perpendicular to the wavevector. Here we instead study propagation parallel to the synthetic magnetic field, in the waveguide geometry illustrated in Fig.~\ref{fig:system} consisting of fluid confined between concentric rotating cylinders. By controlling the cylinder radii and rotation speeds, one can induce either an isolated flux tube or a uniform magnetic field, resulting in effective Schr\"odinger equations previously analyzed in the context of electron vortex beams~\cite{bliokh_electron}. We demonstrate some interesting features and qualitative differences compared to this previous study due to the vectorial nature of the acoustic waves and their differing boundary conditions. 

We find that a uniform synthetic magnetic field can host localized guided modes, which is a nontrivial result due to the constraint that the field must be weak to avoid Taylor instabilities of the background Couette flow~\cite{topo_acoustics,TC_flow}. While the modal energy density is independent of the relative sign of the field and beam orbital angular momentum, the individual pressure and velocity field components are sensitive to this relative sign. As localized beams carrying orbital angular momentum, their longitudinal spin density is also nonzero near the vortex core. A synthetic flux tube does not host localized modes but can nevertheless strongly affect the profiles of non-diffracting Bessel beams. By tuning the strength of the enclosed flux one can tune both the magnitude and sign of the spin density, as well as the beam radius. In both cases, the beam energy density remains finite at zeros of the corresponding Schr\"odinger wavefunction. Near these points the beam's canonical and spin momentum have opposite directions and similar magnitudes, meaning that a weak vector potential can significantly affect the direction of the local energy flow. 

The structure of this article is as follows: Sec.~\ref{sec:review} presents a self-contained overview of the quantum-like formalism for acoustic waves, extending the free space analysis of Ref.~\cite{bliokh2019a} to a background vector potential. Sec.~\ref{sec:wg} analyzes two interesting limits of the acoustic waveguide shown in Fig.~\ref{fig:system}: Laguerre-Gauss beams supported by a uniform synthetic magnetic flux (Sec.~\ref{sec:LL}) and Bessel beams threaded by an Aharonov-Bohm flux tube (Sec.~\ref{sec:AB}). Sec.~\ref{sec:conclusion} concludes with discussion and a summary of the main results.

\section{Vector potentials for acoustic waves}
\label{sec:review}

We study the propagation of small amplitude acoustic waves in a fluid with uniform density and a steady state position-dependent velocity profile $\bs{u}(\bs{r})$. We assume the fluid speed is much smaller than the speed of sound $c$, i.e. $u^2/c^2 \ll 1$, and that $\bs{u}$ has a slow spatial variation compared to the acoustic wavelength. Under these conditions, the acoustic velocity potential $\phi$ obeys the wave equation~\cite{derivation,topo_acoustics}
\be 
\nabla^2 \phi - \frac{1}{c^2} D_t^2 \phi = 0, \quad \quad D_t = \partial_t + \bs{u} \cdot \nabla, \label{eq:second_order}
\ee
where $c^2 = 1/(\rho \beta)$ is the speed of sound, determined by the mass density $\rho$ and compressibility $\beta$. The local pressure $P(\bs{r},t)$ and particle velocity $\bs{v}(\bs{r},t)$ fields induced by the acoustic wave are given by $P = \rho D_t \phi$ and $\bs{v} = - \nabla \phi$. Eliminating $\phi$ yields coupled first order equations~\cite{derivation},
\be
\beta D_t P = -\nabla \cdot \bs{v}, \quad \rho D_t \bs{v} = -\nabla P. \label{eq:first_order} 
\ee
Using Eqs.~\eqref{eq:first_order} and the fact that $\nabla \cdot \bs{u} = 0$ for a time independent background flow, one can show that $P$ and $\bs{v}$ obey the continuity equation
\be 
D_t \left(  \frac{\beta}{2}  P^2 + \frac{\rho}{2} \bs{v}^2 \right) + \nabla \cdot \left( P \bs{v}\right) = 0, \label{eq:continuity} 
\ee
which we will see corresponds to an energy density independent of the background flow, and an energy flux density modified by the background flow. 

Now we specialise to monochromatic waves with time dependence $e^{-i \omega t}$ described by complex $\phi$, $P$, and $\bs{v}$ fields, such that $D_t = -i \omega + \bs{u} \cdot \nabla$. Physical observables are then computed from the real parts of the fields. Assuming $u^2/c^2 \ll 1$, Eq.~\eqref{eq:second_order} can be recast as~\cite{topo_acoustics}
\be 
(\nabla - i \bs{A})^2 \phi + \frac{\omega^2}{c^2} \phi = 0, \quad \bs{A} = -\frac{\omega \bs{u}}{c^2}, \label{eq:scalar}
\ee
such that the background fluid flow $\bs{u}$ resembles a vector potential in the time independent Schr\"odinger equation. The time-averaged energy $W$ and energy flux $\bs{\Pi}$ densities of Eq.~\eqref{eq:continuity} are
\begin{align}
& W = \frac{1}{4} \left( \beta  |P|^2 + \rho |\bs{v}|^2 \right), \\
& \bs{\Pi} = \frac{1}{2} \mathrm{Re}[P^* \bs{v}] + W\bs{u}. \label{eq:kinetic}
\end{align}
Following Ref.~\cite{bliokh2019a}, one can introduce a quantum-like formalism by defining the four-component ``wavefunction'' $\mid \Psi \rangle = (P,\bs{v})^T$ and inner product
\be 
\langle \Psi \mid \Psi \rangle = \frac{1}{4\omega} ( \beta |P|^2 + \rho  |\bs{v}|^2  ),
\ee
such that
$W = \langle \Psi \mid \omega \mid \Psi \rangle$ becomes the local expectation value of the energy operator $\omega$ and $\bs{\Pi}$ is the kinetic momentum density. One can similarly introduce the canonical momentum density $\bs{p} = \langle \Psi \mid (-i\nabla) \mid \Psi \rangle$ characterizing the local phase gradient~\cite{bliokh2019a},
\be 
\bs{p} = \frac{1}{4\omega} \mathrm{Im} [\beta P^* \nabla P + \rho \bs{v}^* \cdot (\nabla) \bs{v} ], \label{eq:canonical}
\ee
where $\bs{v}^* \cdot (\nabla) \bs{v} \equiv \sum_j v_j^* \nabla v_j$. We stress that these quantities generally differ from the energy, canonical momentum, and kinetic momentum densities of the velocity potential, which are $|\phi|^2$, $\mathrm{Im}[\phi^* \nabla \phi]$, and $\mathrm{Im}[\phi^* \nabla \phi] - \bs{A} |\phi|^2$ respectively.

Expanding Eq.~\eqref{eq:kinetic} in terms of the canonical momentum density $\bs{p}$ yields
\be 
\frac{\bs{\Pi}}{c^2} = \bs{p} + \frac{1}{4} \nabla \times \bs{S} - \frac{W}{\omega} \bs{A}, \label{eq:kinetic2}
\ee
where the spin angular momentum density
\be 
\bs{S} = \frac{\rho}{2\omega} \mathrm{Im} \left( \bs{v}^* \times \bs{v}  \right), \label{eq:spin}
\ee
also contributes to $\bs{\Pi}$~\cite{bliokh2019a}. The last term in Eq.~\eqref{eq:kinetic2} can be written as the expectation value of the vector potential,
\be 
\frac{W}{\omega} \bs{A} =  \langle \Psi \mid \bs{A} \mid \Psi \rangle.
\ee
which resembles the vector potential contribution to the kinetic momentum in the Schr\"odinger equation.

For completeness, we also give the angular momentum densities: the canonical angular momentum density,
\be 
\bs{L} = \bs{r} \times \bs{p},
\ee
the total canonical angular momentum density
\be 
\bs{J} = \bs{L} + \bs{S},
\ee
and the kinetic angular momentum density
\be 
\bs{M} = \bs{r} \times \bs{\Pi} / c^2.
\ee 
In particular, it was shown in Ref.~\cite{bliokh2019a} that for localized acoustic fields in the absence of a vector potential, $\langle \bs{S} \rangle = 0$ and $\langle \bs{M} \rangle = \langle \bs{J} \rangle = \langle \bs{L} \rangle$. In the presence of an acoustic vector potential $\bs{L}$ and $\bs{S}$ are unchanged, but $\bs{M}$ is modified by the background fluid flow,
\be 
\bs{M} = \frac{1}{2c^2} \mathrm{Re}[P^* \bs{r} \times \bs{v}]  - \frac{W}{\omega} (\bs{r} \times \bs{A}).
\ee
Interestingly this Schr\"odinger-like formalism for acoustic vector potentials differs from both the scalar Schr\"odinger equation (since $\mid \Psi \rangle$ is a vector field) and the spinor Schr\"odinger-Pauli equation (since no Stern-Gerlach term $\bs{S} \cdot (\nabla \times \bs{A})$ appears in the equations of motion). 

As a simple example illustrating these differences consider the inhomogeneous transverse background flow $\bs{u} = (0,B_z x,0)$, corresponding to a uniform synthetic magnetic field parallel to the $z$ axis. There exists acoustic wave solutions propagating along the $z$ axis and evanescent along the $x$ axis, described by the velocity potential $\phi = e^{i k_z z - \kappa x}$, where $k_z$ and $\kappa$ are the propagation constant and localization length respectively. The pressure and velocity fields are
\be 
P = -i \omega \rho e^{i k_z z - \kappa x}, \quad \bs{v} = \left( \begin{array}{c} \kappa \\ 0 \\ -ik_z \end{array} \right) e^{i k_z z - \kappa x},
\ee
and the dispersion relation is $\omega^2 = c^2 (k_z^2 - \kappa^2)$. This solution is independent of the synthetic magnetic field strength $B_z$, which would couple $v_x$ and $v_y$ if there were a Stern-Gerlach term similar to the Pauli-Schr\"odinger equation. Only the kinetic momentum density is affected by the vector potential,
\be 
\frac{\bs{\Pi}}{W} = \left( 0, B_z x, c \sqrt{1 - \frac{\kappa^2}{k_z^2}} \right),
\ee
acquiring a deflection parallel to the background flow. Since the derivation of the acoustic wave equations assumes a weak background flow $( B_z x / c)^2 \ll 1$, this deflection is small in the plane wave limit $\kappa = 0$. However, as the field becomes more strongly evanescent the longitudinal kinetic momentum $\Pi_z$ becomes smaller, such that $\bs{\Pi}$ tilts more strongly in the direction of the background flow. There are two ways this stronger tilt can be interpreted: (i) the evanescent part of the wavevector contributes to the energy density $W$, but not the first ($\bs{u}$-independent) term of Eq.~\eqref{eq:kinetic}, or (ii) the evanescent wave has a transverse spin $S_y$ which suppresses the longitudinal component of the kinetic momentum in Eq.~\eqref{eq:kinetic2}. The origin of both these effects is ultimately the vectorial nature of the acoustic wavefunction $\mid \Psi \rangle$. By contrast, for scalar (Schr\"odinger) waves with fixed $k_z$ the direction of ${\bs \Pi}$ is independent of $\kappa$.

\section{Acoustic beams in cylindrical waveguides}
\label{sec:wg}

Now we will study the effect of the acoustic vector potential on the propagation of acoustic vortex beams. We shall consider the simplest case of cylindrical symmetry, corresponding to a purely azimuthal background flow $\bs{u} = u_{\theta} (r) \hat{\theta}$, where $(r,\theta,z)$ are cylindrical coordinates. The velocity potential for a cylindrically-symmetric beam with canonical orbital angular momentum $l$ is
\be 
\phi(\bs{r}) = \psi(r) e^{i (k_z z + l \theta)}.
\ee
Substituting $\psi(r)$ into Eq.~\eqref{eq:scalar} yields an equation for the radial profile,
\be 
\left[ \frac{1}{r} \partial_r (r \partial_r) - \frac{1}{r^2} \left(l + \frac{r \omega u_{\theta}}{c^2}\right)^2 + \frac{\omega^2}{c^2}\right] \psi = k_z^2 \psi. \label{eq:radial}
\ee
The pressure $P$ and velocity $\bs{v}$ fields are
\begin{align} 
P &= -i \rho\left(\omega - \frac{u_{\theta} l}{r} \right) \psi e^{i (k_z z + l \theta )}, \\
(v_r,v_{\theta},v_z) &= \left(\partial_r \psi, \frac{il}{r} \psi, i k_z \psi \right) e^{i (k_z z + l \theta) }.
\end{align}
Only $P$ depends explicitly on the vector potential.

To proceed we need to specify the background flow $\bs{u}$. We will consider Couette flow in an incompressible fluid confined between circulating concentric cylinders with radii $R_{1,2}$ rotating at angular frequencies $\Omega_{1,2}$, see Fig.~\ref{fig:system}. The small amplitude acoustic waves obey the hard wall boundary conditions $\hat{r} \cdot \bs{v}(R_{1,2}) = 0$. Assuming the cylinders are sufficiently long that boundary effects at their top and bottom are negligible, the cylinders establish the azimuthal background flow~\cite{TC_flow}
\be 
\bs{u}(r) = \left( \frac{\Omega_2 - \Omega_1\frac{R_1^2}{R_2^2}}{1 - R_1^2 / R_2^2} r + \frac{R_1^2 (\Omega_1 - \Omega_2)}{1 - R_1^2 / R_2^2} \frac{1}{r} \right) \hat{\theta}. \label{eq:u}
\ee
In the following we will focus on two simple limits: a thin magnetic flux line at $r=0$ ($\bs{u}(r) \propto \frac{1}{r} \hat{\theta}$) and a uniform magnetic field ($\bs{u}(r) \propto r \hat{\theta}$). 

At first glance this might seem identical to the electron vortex beam problem studied in Ref.~\cite{bliokh_electron}. However, we stress that there are a few important differences even without taking the vectorial nature of the acoustic waves into account: Acoustic beams obey the Neumann boundary conditions $\partial_r \psi(R_{1,2}) = 0$, whereas an electron beam in a hard wall cylindrical waveguide would have the Dirichlet boundary conditions $\psi(R_{1,2}) = 0$. Additionally, the acoustic vector potential describes a real background fluid flow which much have a finite speed much slower than the fluid speed of sound $c$: the Schr\"odinger-like Eq.~\eqref{eq:scalar} neglects terms of order $(u/c)^2$, and more importantly, the azimuthal flow Eq.~\eqref{eq:u} becomes unstable above a critical fluid speed~\cite{TC_flow,topo_acoustics}. 

\subsection{Uniform synthetic magnetic field}
\label{sec:LL}

We obtain a uniform synthetic magnetic field from Eq.~\eqref{eq:u} by assuming a stationary inner cylinder $\Omega_1=0$ and taking the limit $R_1 \rightarrow 0$, which yields $\bs{u}(r) = r \Omega_2 \hat{\theta}$ and the radial equation
\be 
\left[ \frac{1}{r} \partial_r (r \partial_r) - \frac{1}{r^2} \left(l + \frac{r^2 \omega \Omega_2}{c^2}\right)^2 + \frac{\omega^2}{c^2}\right] \psi = k_z^2 \psi. 
\ee
The well-known solutions are Laguerre-Gauss modes~\cite{bliokh_electron},
\be 
\psi(r) = \left( \frac{r}{w} \right)^{|l|} L_n^{|l|}\left(\frac{2r^2}{w^2}\right) e^{-r^2/w^2}, \label{LGbeam}
\ee
with beam waist $w = \sqrt{2c^2/(\omega |\Omega_2|)}$ and dispersion
\be 
\left( \frac{c k_z}{\omega}\right)^2 =  1 +  \frac{2 l\Omega_2}{\omega} -   \frac{2|\Omega_2|}{\omega} (2 n + |l| + 1), \label{LGdispersion}
\ee
where $n$ is the radial mode number. The mode localization (determined by the beam waist $w$) and energy both depend on the synthetic magnetic field strength, with the dispersion relation including a Zeeman-like interaction between orbital angular momentum $l$ and synthetic magnetic field. This results in an $l$-dependent modal cutoff frequency when $l \Omega_2 < 0$. In contrast, the modal profile $\psi(r)$ is independent of the sign of $l$, even though the vector potential breaks time-reversal symmetry. These features resemble the electron beam case~\cite{electron_review,bliokh_electron}. 

\begin{figure}

\includegraphics[width=0.8\columnwidth]{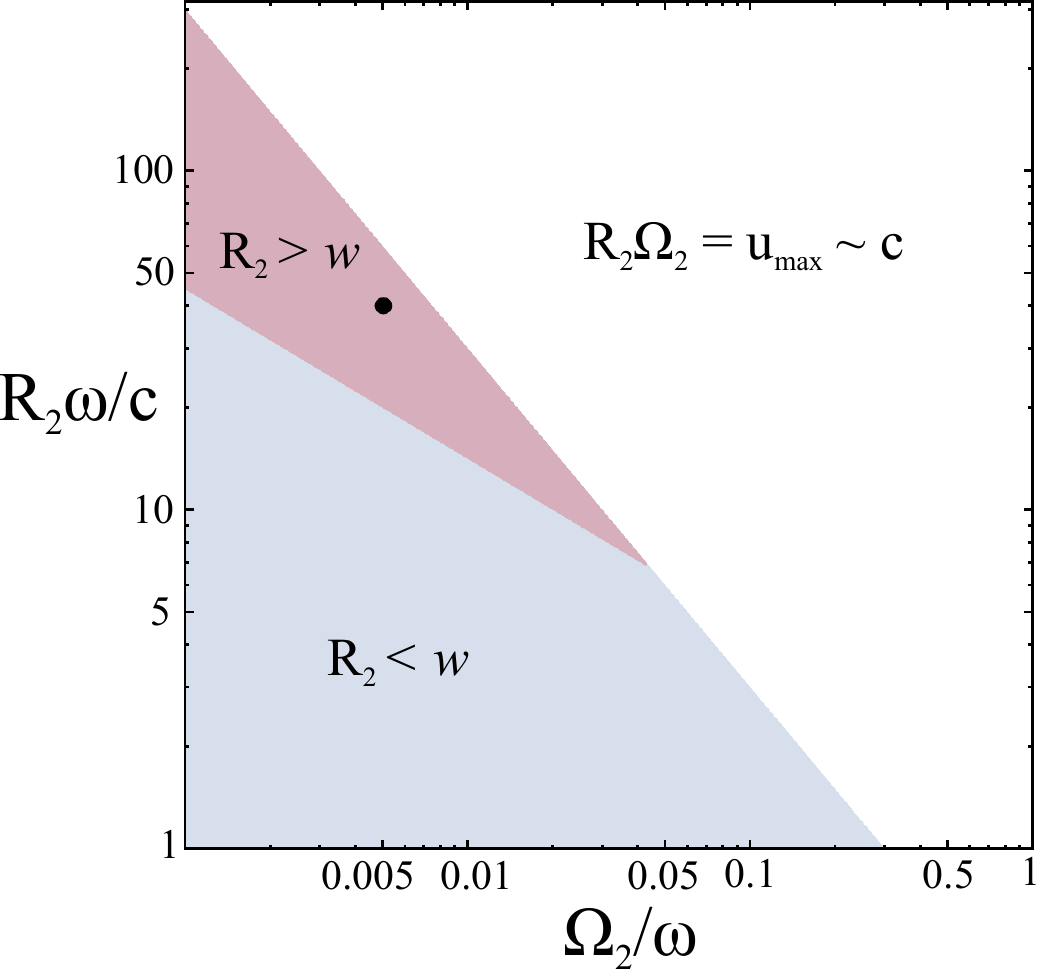}

\caption{Phase diagram of acoustic vortex beams supported by uniform magnetic field induced by rotation $\Omega_2$ of the outer cylinder of radius $R_2$. $\omega$ and $c$ are the acoustic wave frequency and speed, respectively. Unshaded region $u_{\mathrm{max}} > 0.3c$: The large peak fluid velocity $u_{\mathrm{max}}$ renders the effective Schr\"odinger invalid. Red region $R_2 > w$: Vortex beams are localized by the synthetic magnetic field. Blue region $R_2 < w$: Vortex beams are localized by the cylinder wall. Black circle denotes the parameters used in Figs.~\ref{fig:LGdensities} and~\ref{fig:LGAM}.}

\label{fig:phase_diagram}

\end{figure}

The Laguerre-Gauss solution is obtained by assuming an unbounded fluid, so it is only valid if the beam width~$\sim w$ is much smaller than the radius of the outer cylinder $R_2$. However, $R_2$ cannot be too large because the effective Schr\"odinger equation is only valid if the fluid speed is slow compared to the speed of sound, i.e. $(u_{\mathrm{max}}/c)^2 \ll 1$. Moreover, $R_2$ must be remain somewhat larger than the acoustic wavelength, meaning $\Omega_2 / \omega$ must be small. Fig.~\ref{fig:phase_diagram} illustrates these various constraints. 

The dispersion relation Eq.~\eqref{LGdispersion} implies the beams will be close to paraxial, i.e. ${\bf k} \approx k_z \hat{z}$. There is a ``Goldilocks'' zone of sufficiently large (but not too large) $R_2$ supporting acoustic vortex beams localized by the synthetic magnetic field. In the following we use $(\Omega_2, R_2) = (0.005\omega, 40 c/\omega)$, considering vortex beams within this zone. We note that vortex beams solutions still exist for $R_2 \lesssim w$, just they will not take the simple Laguerre-Gauss form Eq.~\eqref{LGbeam}.

\begin{figure}

\includegraphics[width=\columnwidth]{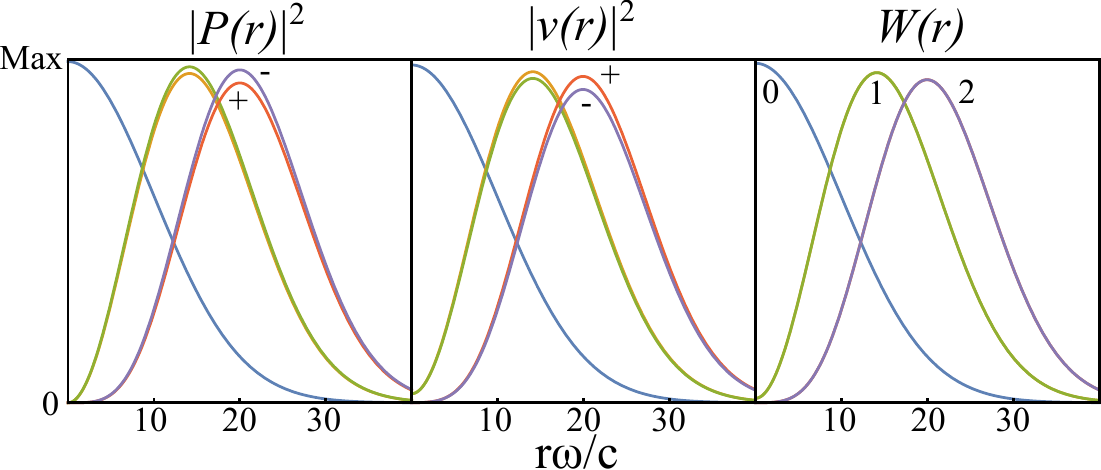}

\caption{Pressure $P$, velocity $|\bs{v}|^2 = |v_x|^2 + |v_y|^2 + |v_z|^2$, and energy density $W$ profiles of Laguerre-Gauss vortex beams of charge $l=0,\pm 1, \pm 2$ confined by a synthetic magnetic field $\Omega_2$. Vortex beams exhibit a field direction-dependent redistribution of energy between their $P$ and $\bs{v}$ components, while $W$ is independent of $\mathrm{sgn}(l \Omega_2)$.}

\label{fig:LGdensities}

\end{figure}

Evaluating the field profiles, there is a $l\Omega_2$-dependent redistribution of energy between the pressure $P$ and longitudinal velocity $v_z$ components,
\begin{align}
|P|^2 &= \rho^2 \omega^2 (1 - \frac{2l\Omega_2}{\omega}) \psi^2, \\
|v_z|^2 &= \frac{\omega^2}{c^2} \psi^2 \left(1 + \frac{2 l \Omega_2}{\omega} - \frac{2 |\Omega_2|}{\omega}(2n + |l| + 1) \right).
\end{align}
However, the total energy density $W$ remains insensitive to $\mathrm{sgn}(l \Omega_2)$. To illustrate these features, Fig.~\ref{fig:LGdensities} plots the energy densities for vortex beams of various charges. Since $\Omega_2 / \omega$ is small there is only a slight (a few percent) redistribution of energy between the $P$ and $v$ fields. Nevertheless, this shows that a detector sensitive to only one of the field components can measure shifts due to the interaction between the beam's orbital angular momentum and the synthetic magnetic field, unobservable for electron vortex beams localized by a magnetic field~\cite{electron_review,bliokh_electron}.

\begin{figure}

\includegraphics[width=\columnwidth]{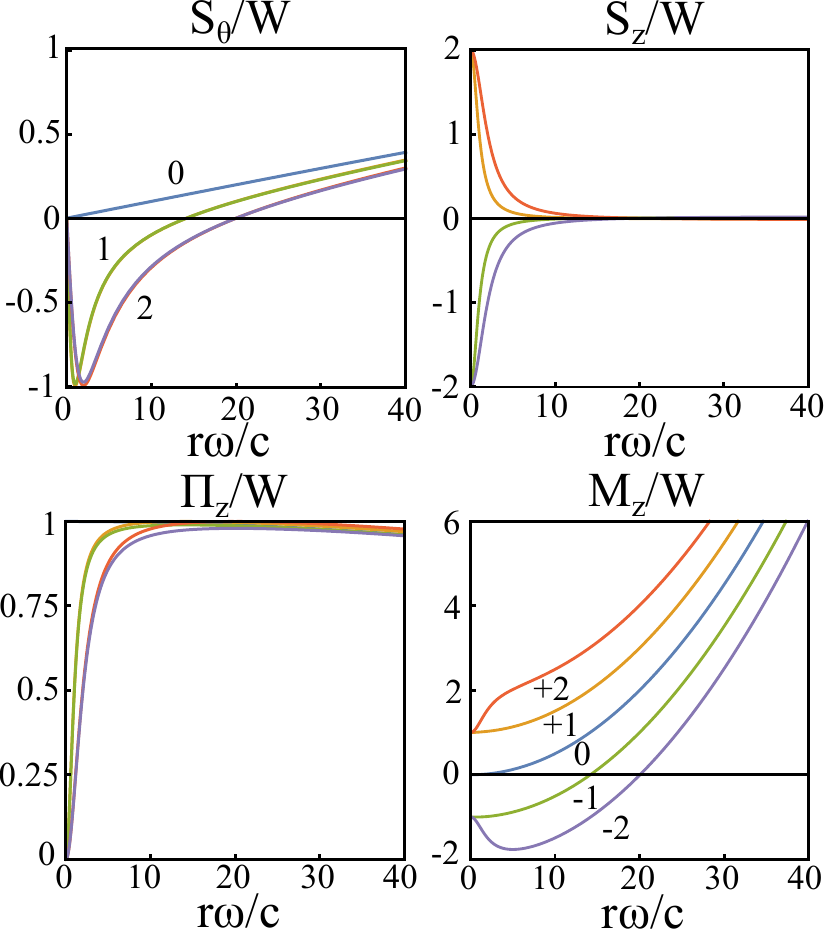}

\caption{Normalised transverse spin $S_{\theta}$, longitudinal spin $S_z$, longitudinal kinetic momentum $\Pi_z$, and kinetic angular momentum $M_z$ densities of the Laguerre-Gauss vortex beams. The synthetic magnetic field does not significantly affect the spin or momentum densities, but induces a change in the sign of $M_z$ for counter-rotating vortex beams with $\mathrm{sgn}(l \Omega_2) < 0$.}
\label{fig:LGAM}

\end{figure}

Since we are close to the paraxial limit, other features of the Laguerre-Gauss beam do not differ significantly from the Schr\"odinger limit. The only exceptions are close to zeros of the potential $\phi$, where non-paraxial corrections can still be significant. Fig.~\ref{fig:LGAM} shows how both the transverse and longitudinal spin are nonzero (comparable to the energy density $W$) close to the core of vortex beams. Consequently the longitudinal kinetic momentum $\Pi_z$ is suppressed by the $\nabla \times \bs{S}$ term in Eq.~\eqref{eq:kinetic2}. Since the background flow is small close to the beam core, there are no obvious synthetic magnetic field-induced changes to the spin densities. On the other hand, the kinetic angular momentum density $M_z$ shows a clear crossover for counter-rotating vortex beams with $\mathrm{sgn}(l \Omega_2) < 0$: for small $r$ the canonical AM (determined by the vortex charge $l$) dominates, whereas at large $r$ the synthetic magnetic field provides the dominates.

\subsection{Aharonov-Bohm flux}
\label{sec:AB}

The second example we consider is a synthetic magnetic flux tube centred at $r=0$. This is obtained by setting $\Omega_2 = 0$ and $R_2 \rightarrow \infty$ in Eq.~\eqref{eq:u}, such that $\bs{u} = \frac{\Omega_1 R_1^2}{r} \hat{\theta}$. Substituting $u_{\theta}$ into the radial equation~\eqref{eq:radial},
\be 
\left[ \frac{1}{r} \partial_r (r \partial_r) - \frac{1}{r^2} \left(l + \frac{\omega \Omega_1 R_1^2}{c^2}\right)^2 + \frac{\omega^2}{c^2}\right] \psi = k_z^2 \psi.
\ee
The general solution decaying to zero as $r \rightarrow \infty$ is~\cite{aharonov}
\be 
\psi(r) = a J_{l + \alpha}(k_{\perp} r) + b J_{-(l + \alpha)}(k_{\perp} r),
\ee
where $J_m(x)$ are Bessel functions, $\alpha \equiv \omega \Omega_1 R_1^2 / c^2$ is the synthetic magnetic flux, and $k_{\perp}^2 = \omega^2 / c^2 - k_z^2$. The hard wall boundary condition $\partial_r \psi(R_1) = 0$ requires
\be 
b = -a \frac{J_{l+\alpha - 1}(k_{\perp} R_1) - J_{l+\alpha+1}(k_{\perp} R_1)}{J_{-l-\alpha - 1}(k_{\perp} R_1) - J_{-l-\alpha+1}(k_{\perp} R_1)}.
\ee
For the special case of integer values of the flux $\alpha$, the Bessel functions $J_{l+\alpha}$ and $J_{-(l+\alpha)}$ become linearly dependent, such that the correct solution is
\be 
\psi(r) = a J_{l+\alpha} (k_{\perp} r) + b Y_{l+\alpha} (k_{\perp} r),
\ee
where $Y_m(x)$ is a Bessel function of the second kind, and 
\be 
b = a \frac{ J_{l+\alpha + 1}(k_{\perp} R_1) - J_{l+\alpha-1}(k_{\perp} R_1)}{Y_{l+\alpha-1}(k_{\perp} R_1) - Y_{l+\alpha+1}(k_{\perp} R_1)}.
\ee
The synthetic magnetic flux $\alpha$ is constrained by $(u_{\mathrm{max}}/c)^2 \ll 1$, which ensures the background flow is slow compared to the acoustic wave speed. The Aharonov-Bohm background flow has its maximum velocity $u_{\mathrm{max}} =\Omega_1 R_1$ at the edge of the inner cylinder, $r=R_1$. Therefore to maximize the flux subject to these constraints a large, slowly rotating inner cylinder is required. In the following we take $\Omega_1 = 0.02$ and $R_1 = 5$, corresponding to a flux $\alpha = 1/2$. 

In contrast to the Laguerre-Gauss beam case, the beam width (determined by the in-plane momentum $k_{\perp}$) is an additional free parameter of the Bessel beam solutions. In the paraxial limit $k_{\perp}$ behaviour similar to Ref.~\cite{bliokh_electron} is expected, since the acoustic wave becomes effectively scalar. On the other hand, in the strongly non-paraxial limit the fluid flow described by the synthetic vector potential is small compared to the large in-plane momentum $k_{\perp}$. Therefore to maximize the effect of the acoustic vector potential, we take a moderately non-paraxial beam angle $\varphi = \pi/8$, where $k_z = k \cos \varphi$ and $k_{\perp} = k \sin \varphi$.

\begin{figure}

\includegraphics[width=\columnwidth]{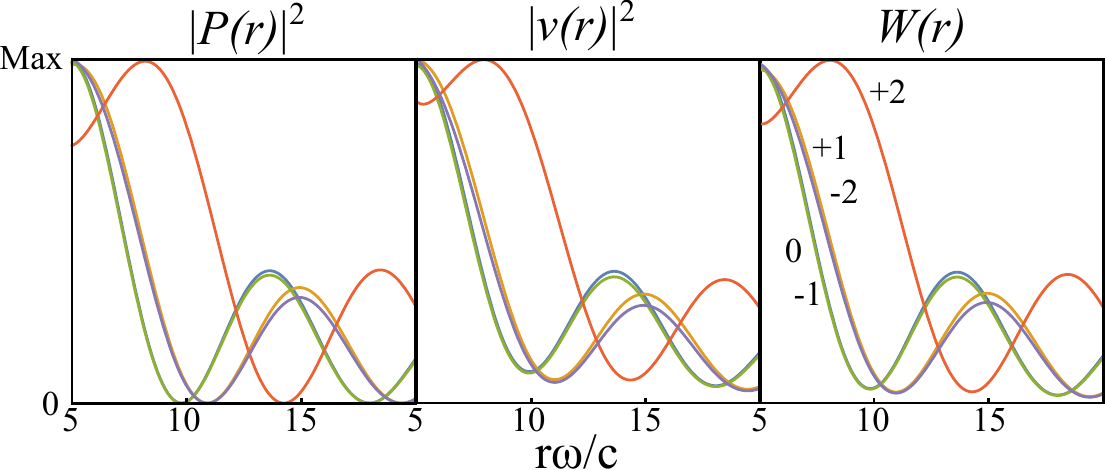}

\caption{Pressure $P$, velocity $|\bs{v}|$, and energy density $W$ profiles of Bessel beams with charges $l=0,\pm 1, \pm 2$ threaded by a synthetic magnetic flux $\alpha$. There is a flux-dependent shift of the minima of $P$, $\bs{v}$, and $W$, with all sensitive to $\mathrm{sgn}(\alpha l)$.}

\label{fig:BBdensities}

\end{figure}

Fig.~\ref{fig:BBdensities} plots $P$, $\bs{v}$, and $W$ profiles of the acoustic Bessel beams. The synthetic flux $\alpha$ strongly breaks the symmetry between beams with orbital angular momentum $\pm l$; the beam radius is larger when $\mathrm{sgn}(\alpha l)>0$. While beams with the same $|l + \alpha|$ have identical velocity potentials $\psi$, they can be still distinguished through their $P$, $\bs{v}$, and $W$ profiles. $P$ vanishes at zeros (corresponding to nodes of $\psi$), while due to non-paraxiality $\bs{v}$ and $W$ are always nonzero. Finally, the inner cylinder boundary $r = R_1$ can be either a global maximum or local minimum of $P$, $\bs{v}$, and $W$, depending on $\alpha$, $k_{\perp}$, and $l$. 

\begin{figure}

\includegraphics[width=\columnwidth]{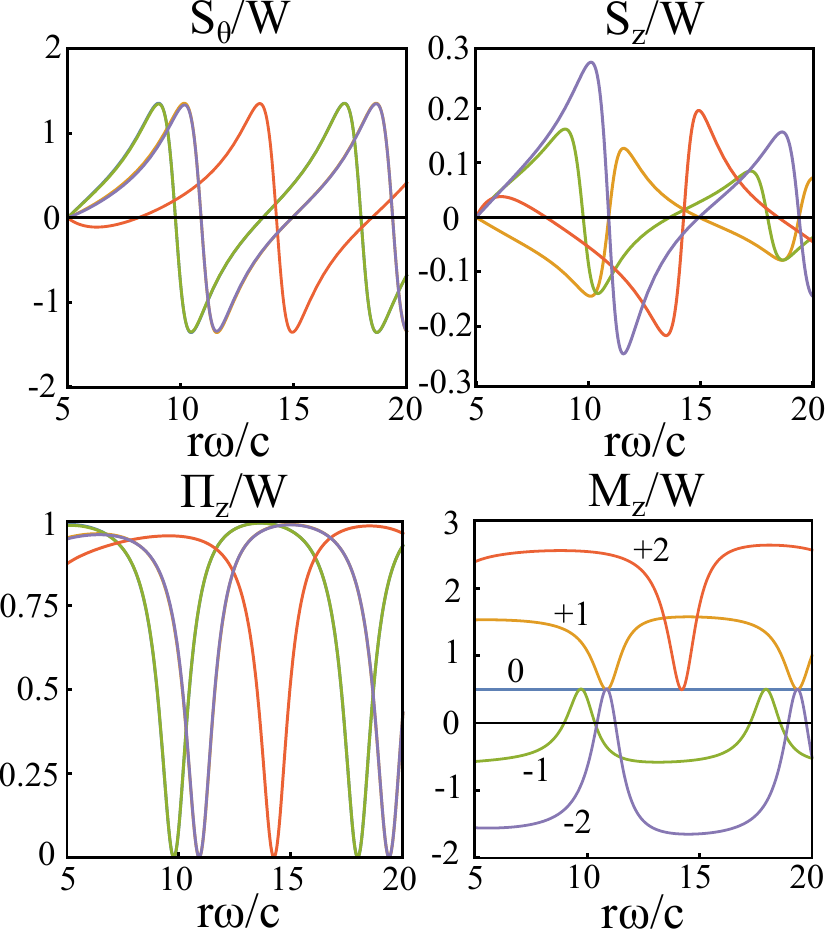}

\caption{Normalised transverse spin $S_{\theta}$, longitudinal spin $S_z$, longitudinal kinetic momentum $\Pi_z$, and kinetic angular momentum $M_z$ densities of the Bessel beams. Shifts in the maxima of $S_{\theta}$ and minima of $\Pi_z$ follow the shifts in the minima of $W$. The peak values of $S_z$ and $M_z$ are sensitive to $\alpha$.}
\label{fig:BBAM}

\end{figure}

Fig.~\ref{fig:BBAM} plot some other observables. The transverse $S_{\theta}$ and longitudinal $S_z$ spin densities become large close to the minima of $W$. In this non-paraxial regime, the peak values of $S_z/W$ are sensitive to the enclosed flux $\alpha$. The longitudinal momentum $\Pi_z$ is determined purely by $|l+\alpha|$, with no visible difference between the $l = 0, -1$ and $l = +1, -2$ beams, and it vanishes at the minima of $W$. Away from the minima of $W$, the kinetic angular momentum $M_z \approx (l + \alpha)W$; at the minima the canonical angular momentum contribution vanishes, leaving $M_z \approx \alpha W$ controlled purely by the enclosed flux.

\section{Conclusion}
\label{sec:conclusion}

In conclusion, we have studied acoustic vortex beams in the presence of synthetic magnetic fields. We analyzed two simple limits; a uniform field and a flux tube, where the vortex beams can be obtained exactly from solutions of the radial Schr\"odinger equation. This enables a direct comparison between previously-studied optical~\cite{bliokh2010} and electron~\cite{bliokh_electron} vortex beams. The main important differences are:
\begin{itemize}
\item The vector potential of the electronic Schr\"odinger equation is a gauge field, whereas the synthetic vector potential in acoustics is the gauge-invariant background fluid speed which must be slow compared to the acoustic wave speed. This limits acoustics to weak uniform synthetic magnetic fields, which can host weakly-localized (near-paraxial) Laguerre-Gauss beams.
\item Beam symmetries present in the effective Schr\"odinger equation and total energy density $W$ can be broken when considering the microscopic acoustic pressure $P$ and velocity $\bs{v}$ fields. This is because the synthetic vector potential induces a redistribution of energy between $P$ and $\bs{v}$.
\item The synthetic magnetic fields can be used to fine-tune the beam profiles, and therefore control the distribution of the acoustic spin density in non-paraxial beams.
\end{itemize}
In the future it would be interesting to extend this analysis to acoustic surface waves, where analogies with surface plasmon polaritons have recently been explored~\cite{Bliokh_arxiv,bliokh2019a,bliokh2019b}. In particular, the breaking of time-reversal symmetry due to the synthetic magnetic field is expected to induce unidirectional surface wave propagation~\cite{bliokh_spp}. Other interesting directions are the analysis of forces on small particles induced by structured acoustic beams~\cite{acoustic_force_bliokh,acoustic_force} and the characterization of random acoustic wave fields, including their distribution of phase and polarization singularities and emergence of superoscillations~\cite{berry2019}. Here we anticipate further non-trivial differences between the scalar effective Schr\"odinger equation governing the acoustic velocity potential and the real measurable pressure and velocity fields.

We thank Konstantin Bliokh for useful discussions. This work was supported by the Institute for Basic Science in Korea (IBS-R024-Y1).

\end{document}